\documentclass{ws-ijmpa}

\begin{document}

\markboth{Costantino Sigismondi}
{Relativistic Implications of Solar Astrometry}

%
\catchline{}{}{}{}{}
%

\title{RELATIVISTIC IMPLICATIONS OF SOLAR ASTROMETRY}

\author{COSTANTINO SIGISMONDI}

\address{Sapienza University of Rome, Physics Dept. P.le Aldo Moro 5 \\
Roma, 00185, Italy\\
University of Nice-Sophia Antipolis - Dept. Fizeau (France); \\
IRSOL, Istituto Ricerche Solari di Locarno (Switzerland)\\
sigismondi@icra.it}

\maketitle

\begin{history}
\received{30 May 2011}
\revised{Day Month Year}
\end{history}

\begin{abstract}
The modern methods of measurement of the solar diameter and oblateness are reviewed.
Either ground-based or balloon-borne and satellite measurements are considered.
The importance of solar astrometry for General Relativity is emphasized, particularly attention is given to the solar oblateness problem,
as well as the studies of solar astrophysics to the whole world of physics from nucleosynthesis to neutrinos.

\keywords{Solar Astrometry; General Relativity; History of Physics.}
\end{abstract}

    
\ccode{PACS numbers: 95.30.Sf, 04.20.-q, 45.50.Pk, 96.12.De, 95.10.Jk, 96.60.-j.}

\section{Introduction}	
The aim of this paper is to show the importance of the field of research of solar astrometry, 
in the framework of solar physics and more generally of all modern physics.
There are several situations in which solar physics is somewhat considered ancient with respect to modern cosmology,
with less top-ranking goals to be achieved, but it is not so.
The lesson of the history shows that the greatest revolutions in the modern physics arose rightly from the field of solar physics.
The comprehension of the nucleosynthesis come from the study of the solar energetics started in the last half of the nineteenth century.
The detection of the anomalous precession of the perihelion of Mercury, firstly attributed to 
a possible solar oblateness, and after to the general relativistic behaviour of the orbit, was understood by Albert Einstein in 1915.\cite{ein} The observation of the total eclipse in Sobral (Brasil) in 1919, steered by sir Arthur Eddington, showed once more the interplay between the Sun and the newborne gravitational physics. 
The researchers continued the measurements of the solar oblateness for the verifications of the theory of General Relativity and for the knowledge of the Sun itself.
Finally in the years 90s of 20th century the solar neutrino problem drove the researchers to the neutrino oscillation which proved the Cabibbo theory of phase mixing, conceived more than thirty years before.
The ultimate tasks on solar physics nowadays deal with the corona mechanism of heating and with the chromosphere, but the so called standard solar model, which is the base of all stellar evolution, is still lacking of the magnetic field which drives the eleven year cycle of the solar spots. 
At present time no solar model is still capable to predict the behaviour of a solar cycle, even after a few years. The current low activity of the Sun since 2007 remains unexplained. The climate studies will benefit of a solar model including magnetic field, for the evaluation of the impact of human activities with respect to the solar activity on the planet Earth for the years to come.

\section{Nucleosynthesis Versus Gravitational Energy}

Several physicists in the nineteenth century believed that the gravitational contraction was the source of energy of the Sun. In 1854 Hermann von Helmholtz proposed this hypothesis. Charles Darwin in 1859 in the first edition of his book on the {\it Origin of species from natural selection} evaluated an age of the Earth and the Sun of about 300 million years. Lord Kelvin, again in 1854, proposed the impact of meteorites as the main source of the energy of the Sun, always a gravitational energy conversion. The chemical source, by chemical reaction, could not provide no more than 3000 years of heat even using all the mass available on the Sun; the gravitational energy could provide 20 million years of energy.
It is of particular interest that the gravitational contraction of the Sun would not be detectable, because of the order of $10^{-6}$ solar radii per year, but the age of the Earth and of the life on the Earth proved the need of a long time duration of solar energetic input.

\noindent The energy obtainable by gravitational contraction is

$\Delta E = \Delta R \times (3GM^2/5R^2)$

\noindent a variation of solar radius of 0.1\% would imply a variation of the energy 75 times the total radiated energy.\cite{Boscardin}
Such a variation in the solar radius could happen only if the mass involved in the process is limited to an external shell.
 
Only with the discovery of Uranium radioactivity by Becquerel and of the heat released by Radio, without cooling, made by Pierre Curie in 1903, drove George Darwin and Ernest Rutherford to propose in 1904 the radioactive disintegration as the real source of solar energy. 
Further observations did not confirm the presence of significant amount of radioactive materials in the Sun.
The theory of Special Relativity of 1905, with the relation mass-energy $E=mc^2$, opened the way to the solution.
Eddington in 1926 wrote {\it The Internal Consitution of the Star} evaluating the internal temperature of the Sun at 40 million degrees, and proposing a simple relation Mass-Luminosity, but the source of the energy was still uncertain between annihilation of proton and electrons or fusion of heavy atoms by single protons. Sir James Jeans was still sustaining the gravitational contraction as the source of solar energy, for his studies on nebulae {\it The Universe around us} of 1929 which paved the way in the studies on cosmology (top-down processes\cite{Sigi01}). The work of Francis W. Aston, Nobel prize for Chemistry in 1922, on the $0.7 \%$ mass defect between Helium and Hydrogen drove Eddington toward the solution of the problem. The calculated lifetime of the Sun become of the order of 100 billion years.
Cecilia Payne in 1925 firstly proved that the Sun was made almost entirely by Hydrogen and George Gamow in 1928 that there is a probability that two charged particle could approach themselves despite of the repulsive potential: the {\it Gamow factor}.
Gamow and Eddington are the fathers of the theory of the proton-proton cycle for nuclear fusion. Carl Friedrich von Weizsaecker discovered the CNO cycle in 1938, were the Carbonium atoms catalize the fusion of Hydrogen. Again in 1938 Hans Bethe published {\it Energy Production in the Stars} which completed the study on nuclear fusion in the stars.\cite{bethe} Gamow included Bethe as second author of the paper written with Alpher, to extend to the cosmological nucleosynthesis the achievements obtained in the stellar physics.\cite{abg}

The works on p-p and CNO cycles contributed to build the current Solar Standard Model. All stellar evolutionary models are based upon this. The Alpher Bethe Gamow theory (1948) on cosmological nucleosynthesis comes from here. 

Nowadays no satisfactory model can yet explain the secular variations of solar activity, nor the difference between the 11-years cycles of solar spots. 

\section{Gravitational Theory: Solar Oblateness and Mercury Perihelion Precession}

The shape of the Sun has been investigated since the anomalous precession of the perihelion of Mercury was found. 

The advancement of the perihelion of Mercury could have been explained by an oblateness of the Sun, the astrometry community performed these measurements finding always new solution to the problem of measuring the solar diameter with an accuracy of one part over 100000, despite of the atmospheric turbulence. The progresses made by Dicke, Hill and Stebbins, Sofia and the SDS team, the RHESSI satellite, the catoptric heliometer of Rio de Janeiro and the Picard french mission are drafted in the following of this section.

The transition between Newtonian and Einstenian gravitation starts in 1859 with Le Verrier's measurements on Mercury's perihelion, completed by Newcomb in 1890s. Einstein published his General Relativity theory in 1915\cite{ein} and Eddington steered the eclipse observational campaign succesfull in Sobral (BR) Island on May 29, 1919 where the gravitational light bending by the solar mass was first observed.

\subsection{Mercury perihelion precession}
According to Le Verrier\cite{flam} and Newcomb's\cite{new,moy} observations (1859-1882) all planetary perturbations yield 
an observed advancement of the perihelion of Mercury's orbit of $574.10\pm0.41$ arcsec per century. 
$42.587\pm0.5$ arcsec/cy remain unexplained by Newtonian theory of gravitation. 
The reference frame for this advancement is also in motion due to the equinox (lunisolar) precession 
(discovered by Ipparchus $\sim$ 150 b.C.) of 50 arcsec per year i.e.~5000 arcsec/cy: it is a motion of the 
Earth's axis i.e. the celestial pole with respect to the ecliptic pole.
Observations from 1765 to now yield an anomalous precession of 
$43.03 \pm 0.03$ arcsec/cy explained by General Relativity as shown in Table I.\cite{scia} 

\begin{table}
\tbl{Planetary perturbations for Mercury. [from Sciama, 1972]}
  {\begin{tabular}{rcl}
    \hline
      Perturbator    & Perturbation [arcsec/cy]  & errorbar [arcsec/cy]   \\
      Venus & 227.856 & 0.27    \\
      Earth & 90.038 & 0.08    \\
      Mars & 2.536 & 0.00    \\
      Jupiter & 153.584 & 0.00    \\
      Saturn & 7.302 & 0.01    \\
      Uranus & 0.141 & 0.00    \\
      Neptune & 0.042 & 0.00    \\
      Solar Oblateness $J_2$& 0.014 & 0.02    \\
      Total Newtonian & 531.513 & 0.30    \\
      Observed (1765-1937) & 574.10 & 0.41    \\
      Difference (obs.) & 42.587 & 0.5    \\
      General Relativity (calc.) & 43.03 & 0.03    \\
    \hline
  \end{tabular}}
\end{table}

\noindent To explain the remaining $42.587 \pm 0.5$ arcsec/cy within Newtonian theory of gravitation have been considered: 

\begin{itemize}
\item{a) the perturbations of an intramercurial planet, Vulcan;}

\item{b) the effects of a small quadrupole moment of the Sun,  yielding a rosette-like orbit with advancing perihelion.}
The perturbation induced by the solar oblateness, assumed as $10^{-7}$ in table I, is very small: 0.014 arcsec/cy.

\item{c) Einstein equations, which fully explain the anomalous precession of the perihelion of Mercury\cite{ein} 
and of the other planets.}

$\delta\theta=6\pi \cdot G M_{\odot} a/c^2b^2$ 

with a,b semiaxes of ellipse $b=a \cdot (1-e^2)^{1/2}$, 
e= eccentricity of the orbit. The observations confirm Einstein's predictions for the advancements of planetary perihelia.
Since   $\delta\theta \propto M_{\odot} /r$, with $r$ orbital distance,  this effect rapidly vanishes for planets far from the Sun.

\end{itemize}

\subsection{Solar oblateness}

Studies on Solar Oblateness were carried by Dicke and Goldenberg:\cite{Dicke1967}
in 1967 their measurements of the solar oblateness have given a value for the fractional difference of equatorial and polar radii $\alpha=(R_{eq\odot}-R_{pol\odot})/R_{eq\odot}$ of $(5.0\pm0.7)\times10^{-5}$, and a corresponding discrepancy of 8\% of the Einstein's value for the perihelion motion of Mercury was implied. From their further analysis of 1974 they found $\alpha=(4.5\pm0.3)\times10^{-5}$ corresponding to $J_2=(2.5\pm0.2)\times10^{-5}$ implying a correction of $(3.0\pm0.3)$ arcsec per century to the classical excess motion of Mercury's perihelion: this through the relation $\alpha=3/2\times J_2$.\cite{Dicke1974} This formula has been upgraded in 1987\cite{Dicke1987} becoming  $J_2=(2/3)(\Delta R_{\odot}-7.8)/R_{\odot}$ where $\Delta R_{\odot}[mas]=R_{eq.\odot}-R_{pol.\odot}$ is the difference between the equatorial and polar solar radii and 7.8 milli arcsec [mas] is the same difference induced only by the surface rotation of the Sun; this part of the oblateness carries no grativational quadrupole moment $J_2$. Hill, Stebbins and Oleson\cite{Hill75} in 1975, refined the acquisition and analysis methods; with the Solar Disk Sextant (1992 – 2009), and  with the satellites RHESSI and Picard (2010-2013) the measurements are carried beyond our turbulent atmosphere. The Newtonian origin of the perihelion precession was progressively ruled out by the more accurate values of the oblateness, cleaned by the effects of the active regions near the limb. 

\noindent There is a Netwonian precession in a quadrupole potential. 
The equation of quadrupole precession is:

\noindent $\Omega_q=-3/2 \bar{\omega}_0 \cdot  (R_{\odot}/r)^2 \cdot cos(i)/(1-e^2)^2  \cdot J_2$,  where

\noindent $J_2=-Q_{33}/2MR_{\odot}^3 $ is the adimensional parameter for quadrupole moment, $R_{\odot}$ the solar radius and $r$ is the orbital semiaxis, $\bar{\omega}_0$ the 
mean motion and $i$ the inclination of the orbit with respect to the equatorial plane.\cite{ciufo}
If $ J_2=10^{-7} $ for the Sun (as from mass, rotation period and solar radius), the contribution to the precession experienced by Mercury should be $0.014$ arcsec/cy, as reported in Table 1.

\subsection{Solar disk sextant}

The Solar disk sextant, SDS, is the most recent experiment on the measurement of the Sun.\cite{sofia,egidi} SDS operated in 1992, 1994, 1995, 1996 and 2009 (data still in analysis).
A rotating telescope above the atmosphere takes the positions of 10 points of the solar disk.
Large photon statistics allow the precise location of those points. 
After data reduction for aberration and optical distortions the expected errorbar on the solar diameter is few milli arcsec. 
the goal of this experiment is to detect secular variations of the solar diameter, beyond the $11$-year sunspots' cycle.

\begin{figure}
\centerline{\psfig{file=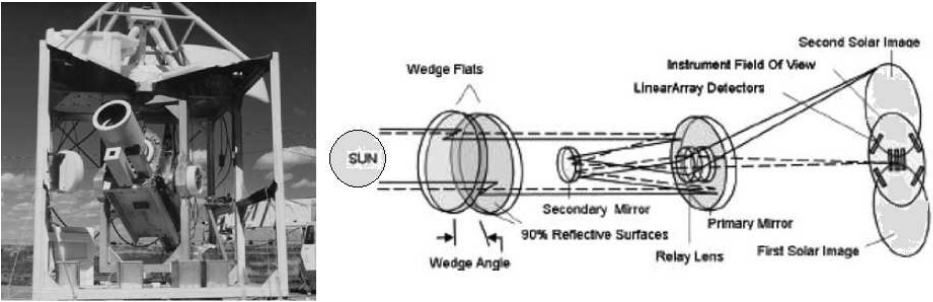,width=13cm}}
\vspace*{0pt}
\caption{Solar Disk Sextant's focal plane linear CCD configuration. The limbs of the two images of the Sun are detected by these linear CCD: 10 points, 5 per each circle, are obtained. After the reconstruction of the two circles, whose distance depends only on the focal length of the telescope and on the beam splitting wedge angle, the solar diameter is obtained from the gap between the two images. This same principle -already used in Goettingen (the same institute built the IRSOL telescope, where now the CLAVIUS project experiments are done, in conjunction with IAP-Paris and ON-Rio de Janeiro) in 1895- is now in the Heliometer of Rio de Janeiro [d'\'Avila {\it et al.}, 2009]. SDS is a Yale-NASA project to which the author has participated.}
\end{figure}

The oblateness measured with SDS\cite{egidi} is plotted in the Fig. 2 among the RHESSI and SOHO/MDI data.

\subsection{RHESSI and Picard}
The shape of the Sun subtly reflects its rotation and internal flows. The surface rotation rate, $\sim 2$ kilometers per second at the equator, predicts an equator-pole radius difference \footnote{This equator-pole radius difference is called also oblateness in some papers, but here we can consider the oblateness as the fractional adimensional parameter $\alpha=(R_{eq\odot}-R_{pol\odot})/R_{eq\odot}$.} of 7.8 milli arcsec, or ~0.001\%. Observations from the Reuven Ramaty High-Energy Solar Spectroscopic Imager (RHESSI) satellite show unexpectedly large flattening, relative to the expectation from surface rotation.  
This excess is dominated by the quadrupole term and gives a total oblateness of $10.77 \pm 0.44$ milli arcsec. The position of the limb correlates with a sensitive extreme ultraviolet proxy, the 284 angstrom limb brightness. The larger radius values are related\cite{rhessi} to magnetic elements in the enhanced network and use the correlation to correct for it as a systematic error term in the oblateness measurement. The corrected oblateness of the nonmagnetic Sun is $8.01 \pm 0.14$ milli arcsec, which is near the value expected from solar rotation. 
The RHESSI measurement essentially follows Dicke's method of using a rapidly rotating telescope to control systematic errors.
\begin{figure}
\centerline{\psfig{file=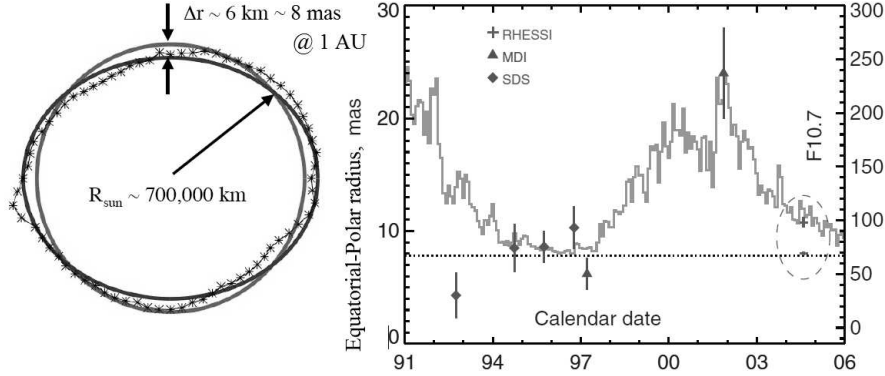,width=12cm}}
\vspace*{0pt}
\caption{Left: solar figure as from the RHESSI measurements. Right: solar oblateness measured by SDS, SOHO/MDI and RHESSI. Evidences of anticorrelation with solar cycle come from SDS data, counter evidences from the combination of all data. This question is still controversial, and the definition of solar limb and of the method of analysis is critical, from [Fivian {\it et al.}, 2008].}
\end{figure}

The French mission Picard (2010-2013)\footnote{http://smsc.cnes.fr/PICARD/}  is designed to measure accurately the parameter $W_{\odot}=dlogR_{\odot}/dlogL_{\odot}$ to recover past solar luminosity (irradiance) $L_{\odot}$ from past values of the radius $R_{\odot}$, obtained from ancient eclipses and so to feed the opportune climate models.
A radiometer SOVAP and the telescope SODISM are performing these measurements.
On the ground a replica of the telescope, SODISM2, will do the measurement in the same time of space, in order to calibrate the method ground-based, which will continue the measurements after the conclusion of the space mission.

\section{Solar Neutrino Defects and the Way for a New Physics}	

About one third of the expected neutrinos from the Sun were not detected. The detectors originally were conceived for the electronic neutrinos and their oscillation in the solar mass and in the space between the Sun and the Earth was proposed as the solution of this lack.
The large experiments of the years 90s of 20th century clarified the neutrino oscillation, and the Cabibbo-Kobayashi-Maskawa model for it. 
The history of the neutrino oscillation is an "Italian brand" either for the measurements (Gran Sasso National Laboratory) and for the interpretation (Nicola Cabibbo, 1963\cite{cabibbo}). The experimental determinations of the solar neutrino flux (Homestake Goldmine 1967, and after GALLEX and SAGE in the Gran Sasso mountain (Italy), Kamiokande and Super-Kamiokande) show a deficit compared to what is predicted by the standard solar model. The solar neutrino defect is attributed to the oscillation of massive neutrinos from one type to another. GALLEX has been the gallium solar neutrino experiments at the Laboratori Nazionali del Gran Sasso from1991 to 1997. The BOREXINO experiment is measuring the 7Be neutrino flux. 
The massive Fermions role in early cosmology,\cite{Sigi01} even if secondary, has been clarified.

\begin{figure}
\centerline{\psfig{file=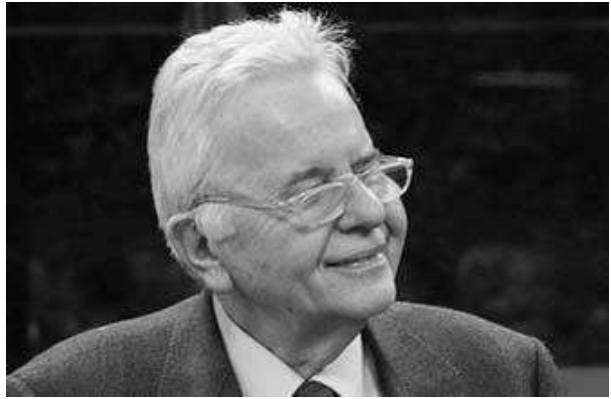,width=8cm}}
\vspace*{0pt}
\caption{Nicola Cabibbo (1935-2010) invented in 1963 the {\it Cabibbo angle}, extended  by Kobayashy and Maskawa in a 3x3 matrix used later to explain the phase mixing of neutrinos. Only the two Japanese received the Nobel prize in 2008: Cabibbo was the president of the Pontifical Academy of Science (composed by lay people and among them 44 Nobel laureates up to now) and the Nobel commission thought right to hit in this way the prestige of that Catholic institution.}
\end{figure}
  
\section{Solar Standard Model: Not Enough to Explain Spots and Secular Cycles}	

The standard solar model has two free parameters: the mixing length scale and the helium abundance, and after 4.52 billion year it should return the present radius of the Sun, its luminosity and the observed metal abundance. During the main sequence phase the solar diameter shrunk of 30\%.
The energy stored in the magnetic field plays a fundamental role in the energy balance of the Sun, as well as the temperature of the photosphere and the diameter of the Sun. The attempt to include the magnetic field in a solar model is still ongoing.
Understanding the reasons of the cyclic variation of the total solar irradiance is one of the most challenging targets of modern astrophysics. These studies prove to be essential also for a more climatologic issue, associated to the global warming. Any attempt to determine the solar components of this phenomenon must include the effects of the magnetic field, whose strength and shape in the solar interior are far from being completely known. Modelling the presence and the effects of a magnetic field requires a 2D approach, since the assumption of radial symmetry -as in the standard solar model- is too limiting for this topic. A 2D evolution code is being developping by the Yale and Rome University solar groups.\cite{ventura} The rotation, the magnetic field and the turbulence are taken into account.

The existence of ice ages either in the last million year and in the pre-Cambrian age has been explained also with astronomical causes (Milankovitch cycles), but the  periods of global warming and cooling in the past millennia have indeed a solar origin. Nowadays the global warming seems to be anthropogenic through the greenhouse effect. But this is still an open question with political, economical and social implications. Conversely there are emerging topics as the global dimming and the debate about the approaching new solar grand minimum.

\subsection{Sun and climate}
The influence of the Sun on the climate of our planet is obvious: all the human activity energetic output is less than 0.01\% of solar input in the delicate climatic system.
For the Earth one hour of exposition to the Sun is equivalent to all energy produced by the human activities in a whole year. 
A predicting model of the solar activity is still lacking either on secular\cite{Gleiss,Usoskin,Yang} and shorter timescales, but also a climatic model based on past solar activity lacks of irradiation data, expected now from past solar eclipses diameters combined with actual $W_{\odot}=dlogR_{\odot}/dlogL_{\odot}$.

\begin{figure}
\centerline{\psfig{file=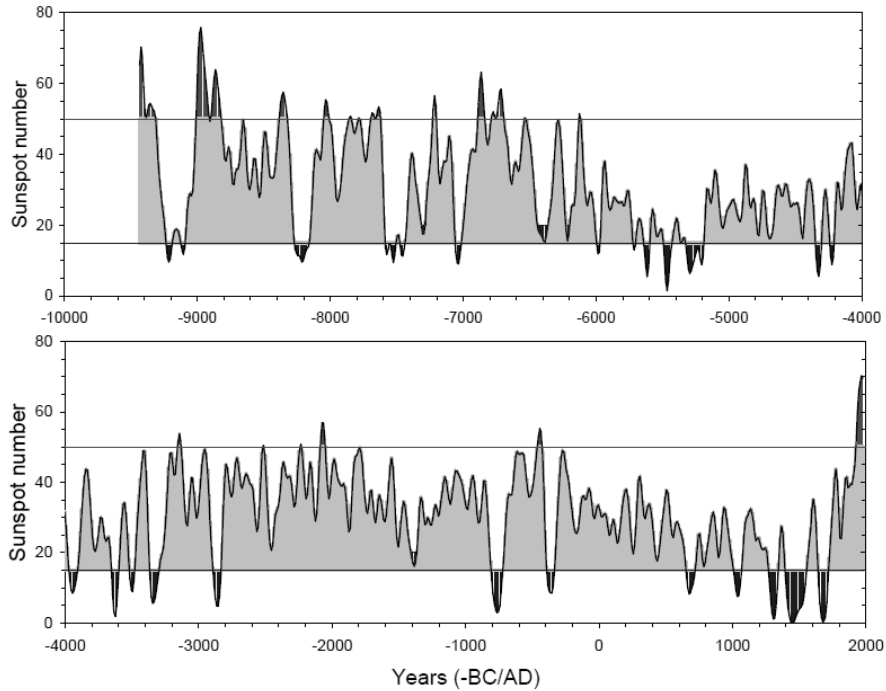,width=12cm}}
\vspace*{0pt}
\caption{Sunspot activity (over decades, smoothed with a procedure introduced by Gleissberg (1944): a 5-point filter is applied with
the consecutive weights 1/8; 2/8; 2/8; 2/8 and 1/8 to each
list of cycle lengths. This filter is usually called the
12221 filter) throughout the Holocene,
reconstructed from 14C [Usoskin {\it et al.}, 2007] using geomagnetic data [Yang {\it et al.}, 2000]. Blue and
red areas denote grand minima and maxima, respectively.}
\end{figure}

\section{The Methods of Measurements of the Solar Diameter}
There are five methods for measuring the solar diameter up to an accuracy of one part over 10000, needed for climatologic studies.
The drift-scan method, adopted since 1655 in Bologna's meridian line built by Giandomenico Cassini and developed in the CLAVIUS 
project;\cite{clavius08} the Heliometer, conceived by Joseph Fraunhofer in 1824 and used also by Wilhelm Bessel to measure the first stellar parallax (1838), the mirror version of that instrument has been invented\cite{helio} at the National Observatory in Rio de Janeiro; the Astrolabe, even in the {\it impersonal} version, conceived by Andr\'e Danjon in 1938 when he was the director of the Paris Observatory and developped in Nice-Calern;\cite{Laclare10} the method of timing the Baily's beads of eclipses introduced by David W. Dunham in 1973;\cite{Dunham73} the transit of Mercury and Venus, used by Irwin I. Shapiro in 1980\cite{shapiro}, with the black drop phenomenon fully understood,\cite{Pasachoff} to evaluate the solar diameter. The balloon and satellite borne experiments exploit the observations from the top of the atmosphere, without seeing turbulences. The figures 5 and 6 are two summarizing tables of this paragraph, while in figure 7 are the puzzling annual averages of daily visual measurements of solar diameters using meridian transits (a drift-scan method), and in figure 8 the also puzzling data of the solar astrolabes worldwide in the last 40 years.
For the eclipses measurements of the solar diameter\cite{guidelines,2006,atlas} also the relativistic corrections to the lunar motion\cite{lunar} have been included. Once considered for the effect of the emission lines of the solar mesosphere,\cite{33} the eclipses provide a long-term data set of reliable values of past solar diameter.\cite{longterm}
The attempt to match eclipses with drift-scan and heliometric measurements is one of the aims of the Clavius project, and the recent mission of the Observatorio Nacional in Rio de Janeiro at Easter Island is among the most promising ones, in terms of the quality of the expected results.
At the Observatorio Nacional in Rio de Janeiro, where the variations of the solar diameter are observed since 1998 with the modified Danjon astrolabe,
also the solar oblateness is now studied with the new catoptric heliometer.

\begin{figure}
\centerline{\psfig{file=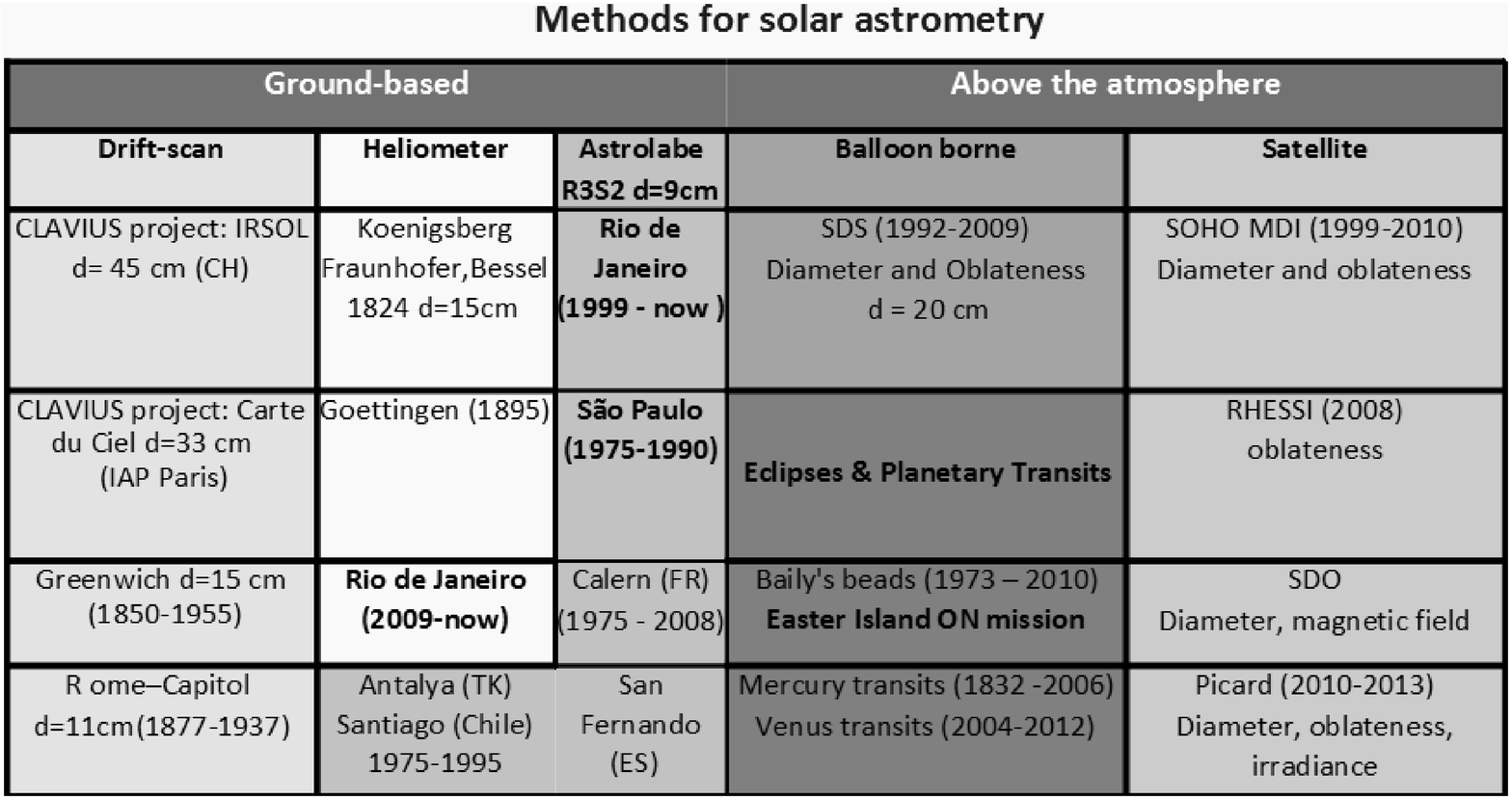,width=12.5cm}}
\vspace*{0pt}
\caption{A table of the methods used to measure the solar diameter.}
\end{figure}

\begin{figure}
\centerline{\psfig{file=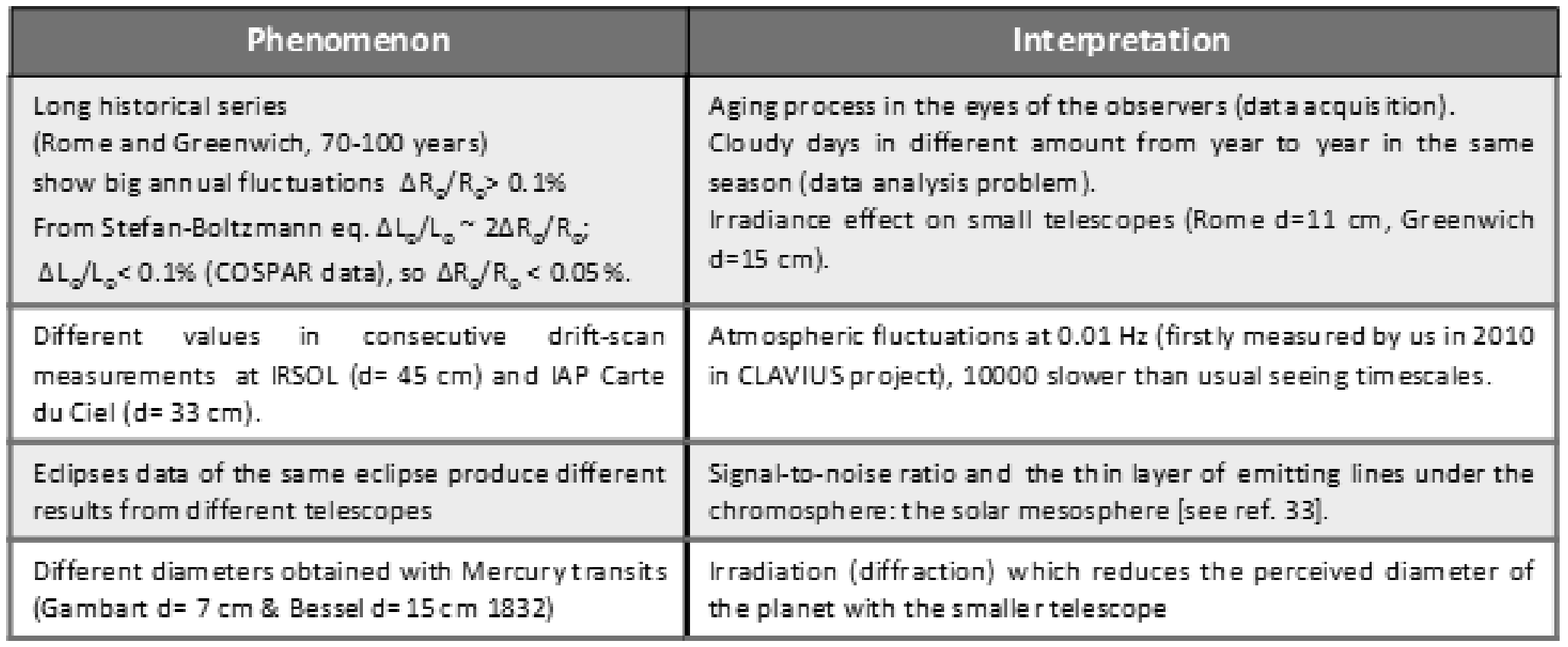,width=12.5cm}}
\vspace*{0pt}
\caption{The problems affecting the accurate measurements of the solar diameter since, at least, two centuries of research. For Mercury transits: the irradiation effect (diffraction); for the drift-scan: the seeing below 0.01 Hz; for the eclipses: solar mesospheric line emission. These are achievements and CLAVIUS project's discoveries.}
\end{figure}

\begin{figure}
\centerline{\psfig{file=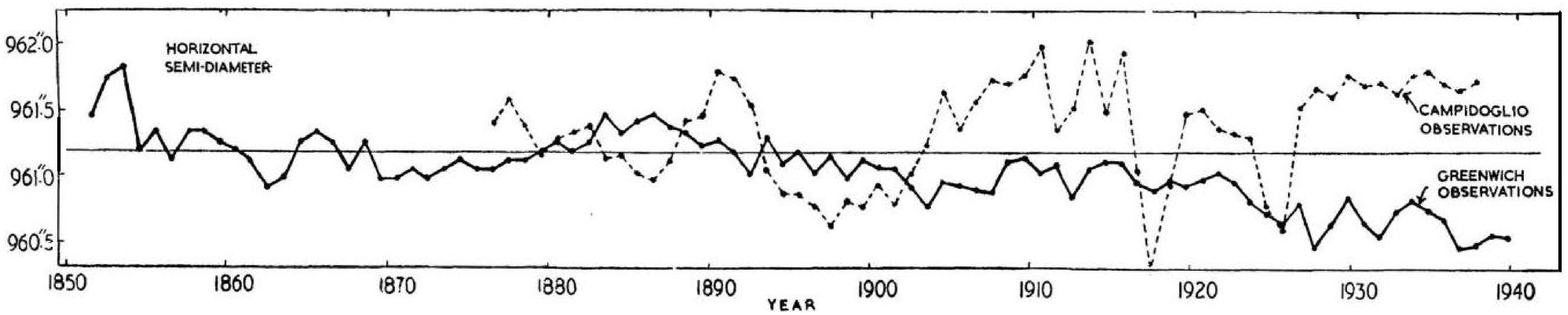,width=13cm,height=4.5cm}}
\vspace*{0pt}
\caption{The meridian transit (drift-scan) measurements operated in Rome Capitol Observatory and in Greenwhich. The large scatters among one yearly average and the other, especially in the roman measurements, are puzzling. From [Gething, 1955]. Recent SOHO data [Bush {\it et al.}, 2010] show no variation beyond $\pm 10$ milli arcsec within 11 years. The standard solar model would not expect $\Delta R_{\odot} > 0.04\%$, unless the shell involved in the change is very thin. }
\end{figure}

\begin{figure}
\centerline{\psfig{file=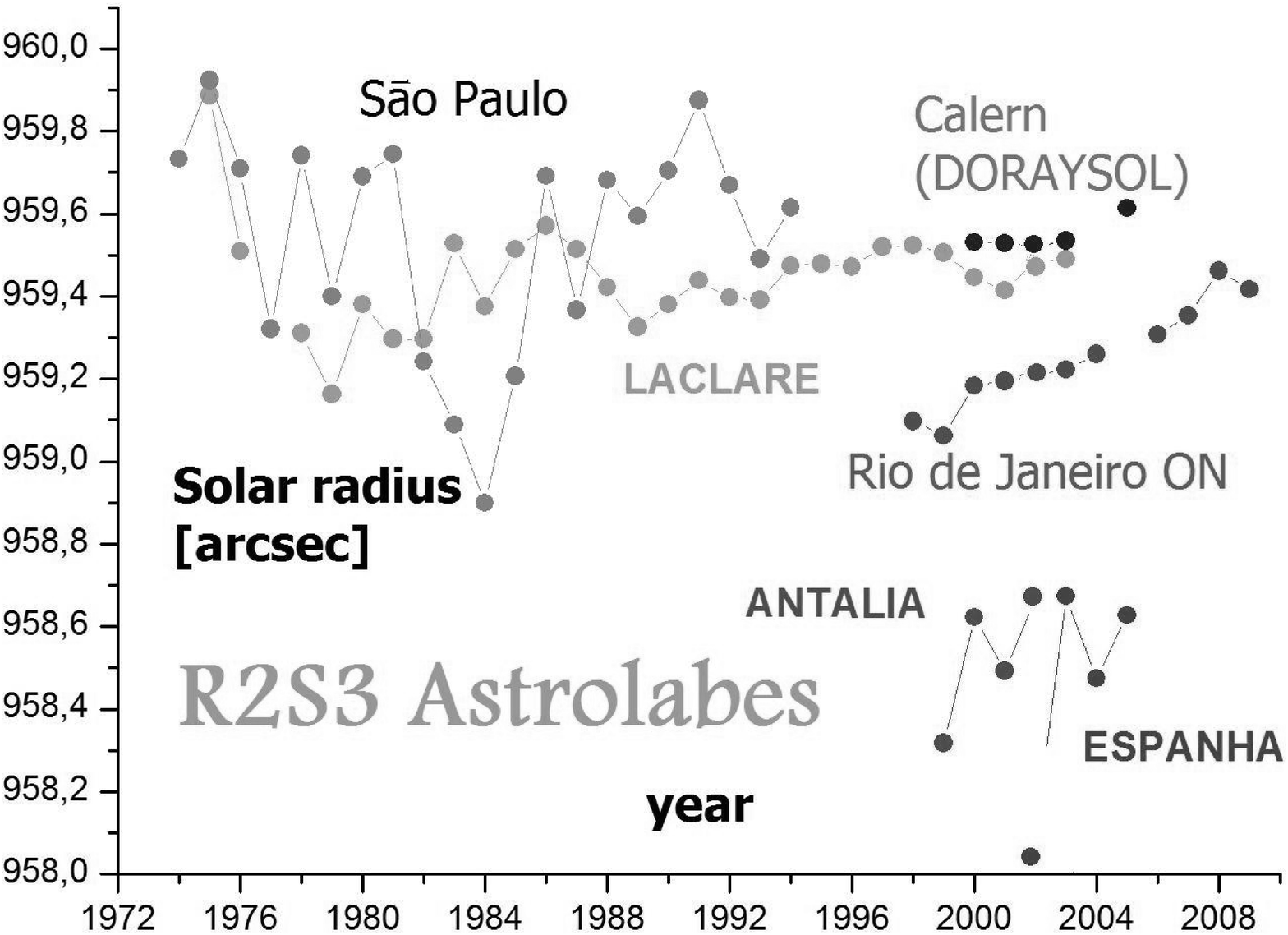,width=13cm}}
\vspace*{0pt}
\caption{Data from various astrolabes of the R\'eseau de Suivi au Sol du Rayon Solaire R2S3 [from S. Boscardin, 2011].
These measurements of the solar diameter are made with astrolabes with slightly different wavebands. This is one reason of the shifts between different sets of measurements. 
They represent the longer series of  available for the last 40 years.
All astrolabes are telescopes with objectives of diameter 10 cm. }
\end{figure}

\section{Conclusions}

The advancements of modern physics in the last 150 years are strictly related with the studies on our closer star: the Sun.
The understanding of nucleosynthesis, also in cosmology; the neutrino phase oscillation and the Einstenian gravitational theory are examples evident to all physicists.
The knowledge of the mechanisms driving and modulating the solar activity over centuries, and even decades, is still incomplete, and the accurate measurements of the solar diameter and of the solar irradiance (once significantly called {\it solar constant}) will help to fill this gap.


\begin{thebibliography}{00}  

\bibitem{ein} A. Einstein,{\it Preuss. Akad. Wiss. Berlin} {\bf 47}, 831 (1915).
\bibitem{Boscardin} S. Calderari Boscardin {\it Um ciclo de medidas do semidiametro solar com astrolabio} (Ph.D. Thesis, Observatorio Nacional, Rio de Janeiro, 2011).
\bibitem{Sigi01} C. Sigismondi, S. Filippi, R. Ruffini and L. A. S\'anchez {\it Int. J. Mod. Phys. D} {\bf 10}, 663 (2001).
\bibitem{bethe} H. Bethe, {\it Physical Review} {\bf 55}, 434 (1938).
\bibitem{abg} R. A. Alpher, H. Bethe, and G. Gamow, {\it Physical Review} {\bf 73}, 803 (1948).
\bibitem{flam} C. Flammarion, {\it Astronomie Populaire} (G. Marpon et E. Flammarion \'{E}diteurs, Paris, 1880).
\bibitem{new} S. Newcomb, {\it Astron. J.} {\bf 14}, 117 (1894). 
\bibitem{moy} A. E. Moyer {\it Simon Newcomb at the Nautical Almanac Office}, in {\it Proc. Nautical Almanac Office Sesquicentennial Symposium} (U.S. Naval Observatory, Washington D. C., 1999), p. 129.
\bibitem{scia} D. W. Sciama, {\it La Relativit\`{a} Generale} (Zanichelli, Bologna, 1972).
\bibitem{Dicke1967} R. H. Dicke and M. Goldenberg, {\it Phys. Rev. Lett.} {\bf 18}, 313 (1967).
\bibitem{Dicke1974} R. H. Dicke and M. Goldenberg, {\it Astrophys. J. Suppl.} {\bf 27}, 131 (1974).
\bibitem{Dicke1987} R. H. Dicke, J. R. Kuhn, and K. G. Libbrecht, {\it Astrophys. J.} {\bf 318}, 451 (1987).
\bibitem{Hill75} H. A. Hill, R. T. Stebbins and J. R. Oleson, {\it Astrophys. J.} {\bf 200}, 484 (1975).
\bibitem{ciufo} I. Ciufolini {\it Physical Review Letters} {\bf 56}, 278 (1986). 
\bibitem{sofia} H. Chiu, E. Maier, K. H. Schatten, S. Sofia, Sabatino, {\it Applied Optics} {\bf 23}, 1230 (1984).
\bibitem{egidi} A. Egidi {\it et al.}, {\it Solar Physics}  {\bf 235}, 407 (2006).
\bibitem{rhessi} M. D. Fivian {\it et al.}, {\it Science} {\bf 322}, 560 (2008). 
\bibitem{cabibbo} N. Cabibbo,  {\it Physical Review Letters} {\bf 10}, 531 (1963). 
\bibitem{ventura} P. Ventura, V. Penza, L. Li, S. Sofia, S. Basu and P. Demarque, {\it Astrophysics and Space Science} {\bf 328}, 295 (2010). 
\bibitem{Gleiss} W. Gleissberg, {\it Terr. Magnetic. Atmos. Electricity} {\bf 49}, 243 (1944).
\bibitem{Usoskin} I. G. Usoskin, S.K. Solanki and G.A. Kovaltsov, {\it Astron. Astrophys.} {\bf 471}, 301 (2007).
\bibitem{Yang} S. Yang, H. Odah and J. Shaw, {\it Geophys. J. Int.} 158 {\bf 140}.
\bibitem{clavius08} C. Sigismondi, M. Bianda and J. Arnaud, {\it Am. Institute of Phys. Proc. Conf.} {\bf 1059}, 189 (2008).
\bibitem{helio} V. d'\'Avila, E. Reis, J. Penna, L. C. Oliveira, A. Coletti, V. Matias, A. Andrei, S. Boscardin, in {\it Solar and Stellar Variability: Impact on Earth and Planets} {\bf IAUS 264} (Cambridge University Press, Cambridge, 2010), p. 487.
\bibitem{Laclare10} F. Morand {\it et al.}, {\it C. Rendus de l'Academie des Science - Physique} {\bf 11}, 660 (2010).
\bibitem{Dunham73} D. W. Dunham and J. B. Dunham, {\it The Moon} {\bf 8}, 546 (1973).
\bibitem{shapiro} I. I. Shapiro, {\it Science} {\bf 208}, 51 (1980).
\bibitem{Pasachoff} J. M. Pasachoff, G. Schneider and L. Golub in {\it Transits of Venus: New Views of the Solar System and Galaxy}, {\bf IAUC 196}, (Cambridge University Press, Cambridge, 2005), p. 242.
\bibitem{guidelines} C. Sigismondi, {\it Science in China G} {\bf 52}, 1773 (2009). 
\bibitem{2006} A. Kilcik, C. Sigismondi, J. P. Rozelot and K. Guhl, {\it Solar Phys.} {\bf 257}, 237 (2009). 
\bibitem{atlas} C. Sigismondi {\it et al.}, {\it Solar Phys.} {\bf 258}, 191 (2009).
\bibitem{lunar} C. Sigismondi, {\it J. Korean Phys. Soc.} {\bf 56}, 1694 (2010). 
\bibitem{33} C. Sigismondi, A. Raponi, C. Bazin and R. Nugent, {\it Int. J. of Mod. Phys. D} submitted (2011).
\bibitem{longterm} C. Sigismondi and A. B. Morcos, {\it Gen. Rel. and Gravitation} {\bf 43}, 1197 (2011).
\bibitem{gething} P. J. D. Gething, {\it Mon. Not. R. Astron. Soc.} {\bf 115}, 558 (1955). 
\bibitem{2010} R. I. Bush, M. Emilio and J. R. Kuhn, {\it Astrophys. J.} {\bf 716}, 1381 (2010).
  
\end{thebibliography}
\end{document}